# Tunability of charge density wave in a magnetic kagome metal

Ji Seop Oh[1,2], Ananya Biswas[2], Mason Klemm[2], Hengxin Tan[3], Makoto Hashimoto[4], Donghui Lu[4], Binghai Yan[3], Pengcheng Dai[2,*], Robert J. Birgeneau[1,5,*], Ming Yi[2,*]

[1]Department of Physics, University of California, Berkeley, California 94720, USA

[2]Department of Physics and Astronomy, Rice University, Houston, Texas 77024, USA

[3]Department of Condensed Matter Physics, Weizmann Institute of Science, Rehovot 7610001, Israel.

[4]Stanford Synchrotron Radiation Lightsource, SLAC National Accelerator Laboratory, Menlo Park, California 94025, USA

[5]Materials Science Division, Lawrence Berkeley National Laboratory, Berkeley, California 94720, USA

*Corresponding authors. E-mail: pdai@rice.edu (P.D.); robertjb@berkeley.edu (R.J.B.); mingyi@rice.edu (M.Y.)

**Abstract**

The discovery of the charge density wave order (CDW) within a magnetically ordered phase in the kagome lattice FeGe has provided a promising platform to investigate intertwined degrees of freedom in kagome lattices. Recently, a method based on post-annealing has been suggested to manipulate the CDW order in kagome FeGe towards either long-range or suppressed orders. Here, we provide a comprehensive comparison of the experimentally measured electronic structures of FeGe crystals that have undergone different post-annealing procedures and demonstrate the remarkable effectiveness on tuning the CDW gap without strong perturbation on the underlying electronic structure. Moreover, we observe an additional low temperature transition that only appears in crystals with a long-range CDW order, which we associate with a lattice-spin coupled order. Our work indicates a likely strong sensitivity of the CDW order to disorder in FeGe, and provides evidence for strong coupling between the electronic, lattice, and spin degrees of freedom in this kagome magnet.

Quantum materials exhibiting comparable energy scales of electron kinetic energy and interactions between quasiparticles hold promise for observing emergent phenomena[1]. In such systems, the presence of electronic instabilities near the Fermi level ($E_F$) can further lead to an abundance of exotic phases. These electronic instabilities not only introduce emergent orders but could also interact with bosonic modes such as phonons and magnons, giving rise to competing or intertwined orders[2–4].

Recent examples of these scenarios are seen in metallic kagome lattices[5–20], where the chemical potential resides near saddle points in the electronic structures, or Van Hove singularities (VHSs), at the boundary of their Brillouin zone (BZ) in momentum space. Theoretical models, considering intra- and inter-site Coulomb interactions, predict distinctive CDW phases characterized by non-zero angular momentum of a particle-hole pair[7,10]. This has been further suggested to indicate the existence of orbital loop currents and nematicity[10,21], differentiating the CDW order in this material from ordinary charge ordering.

One system that embodies this prediction is the vanadium-based kagome lattice, $A$V$_3$Sb$_5$ ($A$ = Cs, K, Rb)[8–20]. Diverse emergent phases have been discovered, including CDW order potentially linked to broken time-reversal symmetry[16] and the presence of two-dome superconductivity[9]. Amidst extensive theoretical modeling and experimental investigations, studies of the electronic structure have identified multiple VHSs near $E_F$. These VHSs exhibit various characteristics, such as normal and higher-order dispersions across the BZ boundary[14] when coupled with CDW. Furthermore, observations of momentum-independent electron-phonon coupling[18] and reports of nodeless superconducting gaps[17,19] underscore the significance of understanding the coupling between the electronic and other degrees of freedom for comprehending emergent orders in kagome metals.

Mn- or Fe-based metallic kagome magnets introduce the magnetic degree of freedom to the emergent kagome physics[22]. FeGe, an A-type antiferromagnet whose onset temperature is at 410 K[23,24], stands out as a special example among the kagome magnets. Structurally, it's been identified as a fundamental building block of these kagome magnets based on an advanced kagome effective Hamiltonian using a $d$-orbital basis[25]. It is also known due to its unique CDW phase below 110 K ($T_{CDW}$)[26–32]. The CDW order in FeGe has been discovered and characterized by neutron scattering, scanning tunneling microscopy, and angle-resolved photoemission spectroscopy (ARPES)[26]. Notably, magnetism in FeGe is expected to be intertwined with the CDW order as the size of the magnetic moment shows an increase across the CDW transition. Furthermore, theoretical frameworks have been established to explain the exotic ground state of FeGe in broader contexts. In the magnetic phase diagram based on a Hartree-Fock approach, FeGe is located at the boundary of spatial stripe spin-charge order[29]. Consequently, FeGe has garnered substantial attention, offering a unique platform to study interplay between the CDW order and magnetism.

In-depth characterizations further elucidate complex coupling between quasiparticles in FeGe. After the initial discovery of the CDW order, an ARPES study uncovered details of the electronic structures including the shift of the VHSs by the magnetism-induced exchange splitting and the momentum-dependent opening of the CDW gap on the VHS-forming band dispersion[28]. In addition, moderate electron-boson coupling was revealed from the same VHS dispersion that has a CDW gap, with an energy scale matching that of an optical phonon mode at approximately 30 meV. A separate study employing elastic and inelastic x-ray scattering observed a large response in the phonon dispersions across both $T_{CDW}$ and the Néel temperature ($T_N$), further demonstrating the intertwinement of charge, lattice, and magnetism[31]. Delicate interactions between coexisting orders were emphasized again by temperature-dependent neutron Larmor diffraction and Raman spectroscopy, revealing a rare symmetry ascension when the temperature is lowered across $T^* = 60$ K[32]. Optical spectroscopy studies reported spectral weight transfer across $T_{CDW}$ at both a small energy scale (< 250 cm$^{-1}$) and a larger scale (500–10000 cm$^{-1}$), speculated to be due to a minuscule Fe displacement and an abrupt shift of VHS, respectively[33,34].

Recently, an annealing method has been reported for manipulating the CDW order in FeGe, which either optimizes it into a long-range order state, in which the correlation length is enhanced 10 times from 2-4 nm, or completely suppresses it, thus providing an opportunity to explore the CDW order and its relations with other degrees of freedom[35–38]. Initially, the CDW order in FeGe was identified as a short-range order in as-grown crystals, and this was confirmed by the line width of the CDW superlattice peaks in neutron diffraction[26]. A study with scanning tunneling microscopy and x-ray diffraction on both the as-grown and post-annealed samples observed and refined long-range 2 × 2 × 2 superstructures[35]. Controlled annealing studies found that annealing at 320 °C for over 48 hours optimized the CDW order into a long-range order while annealing at 560 °C completely suppressed it[36,37]. It is therefore imperative to probe experimentally how the electronic structures in FeGe are affected by the annealing conditions as the CDW order is tuned, and ultimately to provide concrete understanding for interactions between the charge, lattice, and magnetism in FeGe.

Here, we present and compare the electronic structures of FeGe annealed under 320 °C (FG320) and 560 °C (FG560) below and above $T_{CDW}$, providing clear signatures from the CDW order. The spectral intensity at $E_F$ near the M point is suppressed only in the CDW phase, consistent with CDW gap formation on the VHS-forming band dispersion. Further measurements of the systematic temperature dependence of the VHS dispersion show that the CDW gap only forms in FG320 below 110 K. In contrast, FG560 shows a much lower onset temperature of such spectral weight suppression, suggesting that short-range CDW fluctuations may persist. In addition, from the detailed temperature evolution of the band structure on both crystals, we identify a band shift at a characteristic temperature scale $T^* < T_{CDW}$ in FG320 only, where the band shift direction is in reverse as those at $T_{CDW}$. In comparison to the known low temperature transitions observed by other experimental probes, we conclude that this band shift is likely a reflection of a change in the lattice, which is coupled to the spin degree of freedom, consistent with reported moderate electron-phonon coupling and optical phonon hardening across $T_{CDW}$.

As shown in Fig. 1A, the crystal structure of kagome FeGe consists of alternating Fe kagome layer and Ge honeycomb layer along the $c$ axis. In accordance with previous reports, we have annealed our crystals at either 320 °C or 560 °C for 96 hours for inducing long-range CDW order in the former and suppressing CDW order in the latter, which we label as FG320 and FG560, respectively. The effects on the phases in FG320 and FG560 are presented in Fig. 1B, consistent with previous reports[35,36]. Both FG320 and FG560 have similar $T_N$ of 410 K. $T^*$, another temperature scale that we will present later, is also indicated for FG320C. Our magnetic susceptibility measurements on the two types of crystals indeed show the sharp CDW order in FG320 and the lack of such anomaly in FG560 (Fig. 1C and 1D).

Now we compare the electronic structures of FG320 and FG560 as measured by ARPES. While FeGe undergoes the lifting of $C_3$ symmetry below $T_N$[32], the quantitative distortion is tiny, hence a hexagonal BZ description is still a valid approximation (Fig. 2A). Since the valence band is dominated by the Fe $3d$ orbitals forming a layered kagome network, we use the surface BZ notations for simplicity. An empirical schematic of the valence band is presented in Fig. 2B, where we use the notations of VHS1, VHS2 to label the VHSs of dominantly $d_{x^2-y^2}/d_{xy}$ orbitals and $d_{xz}/d_{yz}$ orbitals, respectively, in accordance with previous work[28].

We first directly compare the measured Fermi surfaces (FSs) of FG320 and FG560 both above and below $T_{CDW}$ (Fig. 2C). The top half of Fig. 2C shows the FSs of FG320 while the bottom half shows the FSs of FG560. The left half of Fig. 2C shows the FSs measured below $T_{CDW}$ at 80 K on the two samples while the right half shows the measurement at 150 K above $T_{CDW}$. All measurements were carried out under identical photon energy and polarization. First, at 150 K, the FSs of FG320 and FG560 appear to be very similar. Out of the four quadrants, only the top left quadrant is obtained in a phase with CDW order. Comparing this data with the others, we identify two main distinctions. First, the spectral intensity between $\bar{K}$ and $\bar{M}$ (marked by an orange rectangle) is suppressed in the CDW state while it is noticeable in all other states. This is consistent with the suppression of spectral weight due to the opening of a CDW gap on the VHSs at $\bar{M}$. Second, we observe a change in the appearance of the circular FS near the $\bar{\Gamma}$ point. This is more apparent in the second BZ ($\bar{\Gamma}_1$) due to less suppression by photoemission matrix elements. Noticeably, the circular pocket is separated from the broad spectral intensity only in the CDW phase. We infer that it may come from orbital-dependent band shifts. Note that the FSs in the non-CDW phase are invariant under different post-annealing conditions, implying that the suppression of the CDW order is intrinsic and not due to sample quality variation.

Next, we present a side-by-side comparison of band dispersions measured along the $\bar{K}$-$\bar{M}$-$\bar{K}$ high symmetry cuts. Figures 2D and 2E show raw ARPES intensity plots at 80 K and 150 K from FG320 (left half in each plot) and FG560 (right half in each plot). Figures 2F and 2G show the same data but overlaid with the corresponding momentum distribution curves (MDCs) at $E_F$ and guidelines using annotations from Fig. 2B. Again, very little difference is observed on the band dispersions of the two samples at 150 K. Comparing the 80 K data, we note that more band dispersions are observed to cross $E_F$ in the CDW phase in FG320, as expected from folding due to the CDW order. This is further supported by peaks found in the MDCs (orange bars) in the CDW phase in addition to those found in the non-CDW phase (green bars) in Fig. 2F. These additional band dispersions are also shown by dashed lines and orange arrows. Note that orange dashed dispersions between $\bar{\Gamma}_1$ and $\bar{K}$ are observed only in the CDW phase.

We further differentiate FG320 and FG560 more directly by examining the CDW gap as a function of temperature. We first discuss our observations on FG320. Figure 3A shows a side-by-side comparison of the dispersions along the $\overline{\text{K}}$-$\overline{\text{M}}$-$\overline{\text{K}}$ taken at 15 K and 150 K. Near $E_F$, we note that the spectral weight is suppressed between $\overline{\text{K}}$ and $\overline{\text{M}}$ in the CDW phase as shown in the left panel (15 K) compared to that on the right panel (150 K), consistent with the observation on the FS maps in Fig. 2C.

More details can be observed by investigating the temperature dependence of the energy distribution curves (EDCs) at two different Fermi momenta, $k_{VHS}$ and $k_0$, defined as the momenta where the bands corresponding to VHS1 and VHS3 cross $E_F$. The Fermi-Dirac (FD) distribution function has been divided from all the EDCs. As the location of both $k_0$ and $k_{VHS}$ vary with temperature (to be discussed in the following section), the EDC at each temperature is taken after precisely determining the momentum location of the band crossing.

The temperature dependence of FD-divided EDCs at $k_0$ and $k_{VHS}$ from FG320 are shown in Fig. 3B and 3C, respectively, with clear indications of a gap opening at $k_{VHS}$ and its closure across $T_{CDW}$ (110 K). In the EDCs taken at $k = k_{VHS}$ (Fig. 3B) below 110 K, one can identify changes in the slopes near $E_F$ and the presence of dips in the intensity, and both are signatures of a finite energy gap. Black arrows indicate locations where slopes change, which is a measure of the gap size. As temperature increases, the gap size shrinks and becomes zero at around 110 K, $T_{CDW}$. The overall shapes of the gapped EDCs agree with that observed on as-grown crystals[28], with a 10K increase in $T_{CDW}$ together with a longer correlation length of the CDW order. In contrast, the EDCs at $k = k_0$ (Fig. 3C) have peaks at $E_F$. The peaks are sharper at lower temperatures, which is the typical behavior of a quasiparticle peak, demonstrating the absence of a CDW gap at $k_0$. Both observations of the presence and absence of the CDW gap at $k_{VHS}$ and $k_0$, respectively, are consistent with momentum-dependent CDW gap opening characterized in the as-grown FeGe. The temperature evolution of the CDW gap in FG320 can be more directly visualized by combining the symmetrized EDC at $k_{VHS}$ into a spectral image (Fig. 3D). Here, the suppressed intensity inside the CDW gap is visible as the white region at $E_F$, closing around $T_{CDW} = 110$ K. We note that the CDW gap is only observed to open on selected bands with a gap size of 20 meV, while most band crossings remain gapless. Hence it remains to be seen the expected effect of such a momentum-dependent gap by optical spectroscopy in the corresponding energy range, which is from the joint density of states over the entire BZ[33,34].

Via the same method, we can examine the temperature evolution of the possible CDW gap in FG560, where no clear long-range CDW characteristic is observed in the magnetic susceptibility measurement. Contrary to FG320 (Fig. 3A), the band dispersions at 15 K and 150 K in FG560 show no detectable differences (Fig. 3E). The spectral intensity at $E_F$ between $\bar{K}$ and $\bar{M}$ remains unchanged in both low and high temperatures. FD-divided EDCs at $k_{VHS}$ and $k_0$ shown in Figs. 3F and 3G provide noticeable differences from those in FG320 (Figs. 3B and 3C). Signatures of the CDW gaps in Fig. 3F are almost absent except for small dip-like features below 35 K. All the EDCs shown in Fig. 3G show peaks at $E_F$ and similar temperature dependence with that shown in Fig. 3C. In the symmetrized EDC spectral image (Fig. 3H), we identify a smaller white region than that for FG320 (Fig. 3D), indicating much weaker CDW nature in FG560. We propose that this weak spectral weight suppression for a limited temperature window may be from fluctuating short-range CDW order. A detailed comparison of the EDCs for the two types of samples is also provided in the SM.

Now we examine the temperature dependence of the electronic structures of the FG320 and FG560 aside from the CDW gap. The detailed temperature-dependent dispersions measured along the $\bar{K}$-$\bar{M}$-$\bar{K}$ high symmetry cut across both $T^*$ and $T_{CDW}$ are shown for FG320 (Fig. 4A). As marked by the guidelines, we identify the CDW gap formation on the VHS1-forming band below $T_{CDW}$, back-bending of the VHS2-forming band and its shift across $T^*$ and $T_{CDW}$. A summary of the band evolution for FG320 across $T^*$ and $T_{CDW}$ is shown by the schematic in Fig. 4B. These evolutions across the transitions can be better visualized by the spectral images made from the temperature-dependence of the MDC taken at $E_F$ (Fig. 4C) and the EDC taken at the $\bar{K}$ point (Fig. 4D). In the MDC spectral image (Fig. 4C), two features are evident between $\bar{M}$ and $\bar{K}$ as associated with VHS1 and VHS3, which are the $k_{VHS}$ and $k_0$ that we used to confirm the CDW gap in Fig. 3. $k_0$ is the momentum where the Dirac band at $\bar{K}$ crosses $E_F$. As temperature is lowered across $T_{CDW}$, $k_0$ shifts towards $\bar{K}$, consistent with a up-shift of this band in energy, as also seen by the temperature-dependent EDC image (Fig. 4D). Interestingly, this shift is reversed when temperature is lowered across $T^*$. In a previous ARPES report on an annealed FeGe, this shift is observed across $T_{CDW}$ and was interpreted to be a manifestation of the Ge dimerization[38]. In addition, a slow energy shift with small slope as a function of temperature is observed, which is due to temperature evolution of spin exchange splitting as reported previously[28].

In comparison, we show the exact same measurements on FG560 (Fig. 4F-J). In contrast to the two scales ($T^*$ and $T_{CDW}$) in FG320 where the $k_0$ shifts, the FG560 sample does not show this band shift at either temperature scale. The dispersions along the $\overline{K}$-$\overline{M}$-$\overline{K}$ at different temperatures are shown in Fig. 4J, with a summary of the observed changes illustrated by a schematic in Fig. 4F. VHS2 has only one band near $E_F$, implying that back-bending and doubling of VHS2 are absent in FG560. The VHS2-forming band dispersion becomes broad and fade above $T^*$ in FG560, while it is noticeable in FG320 in the same temperature range. From this set of temperature dependent measurements, we can again extract the temperature evolution of the MDC at $E_F$ (Fig. 4G) and EDC at $\overline{K}$ (Fig. 4H). Most evidently, the MDC shows no change across $T_{CDW}$ nor $T^*$, not even broadening of the bands, indicating that the lack of CDW order here is not due to mixed regions with CDW and regions without CDW, nor due to low spectra quality from a bad cleavage, but a demonstration of a lack of ordering. In addition, we note that change of VHS3 (grey guidelines) are not as significant as in FG320 in which two non-monotonic shifts are captured across $T^*$ and $T_{CDW}$, as shown in Fig. 4H.

By carrying out systematic and high quality ARPES measurements under identical conditions, we have presented a systematic comparison and contrast of the electronic structure of FG320 and FG560, which are kagome FeGe post-annealed at 320 °C and 560 °C, respectively. We summarize our observations organized by three temperature ranges separated by $T_{CDW}$ and $T^*$. First, above $T_{CDW}$, both types of crystals are in the A-type antiferromagnetic order. Here we see that the electronic structures of the two crystals are largely identical, with nearly no observable differences. This can be best seen in the FS comparison in Fig. 2B and the dispersion comparison in Fig. 2D for 150 K. Comparing the peak width in the MDC at $E_F$, the peaks are slightly broader in FG560, indicating that the subtle difference may be due to a slightly larger amount of disorder in FG560, as suggested from a previous report[35,37]. In addition, the nearly identical location of the VHSs in the two types of crystals indicate that the VHSs being near $E_F$ by itself is not a sufficient condition for driving the CDW order. The amount of disorder may also play a role. Here we would like to note that a previous ARPES work concluded that FS nesting does not exist in FeGe based on the auto-correlation of a single FS mapped under one photon polarization [38]. We would like to caution that the photoemission matrix elements associated with a particular polarization suppress intensity from orbitals of certain symmetries. Hence the electronic structure including the FS measured under one particular polarization is incomplete. As has already been demonstrated by previous work[28], the FSs measured under different polarizations reveal different VHS-associated bands. Hence the analysis for nesting based on auto-correlation of the measured FS strongly depends on the polarization and is only conclusive when all features are observed. With that said, what we observe here is that even though the VHSs are near the $E_F$ here in both types of annealed samples, they are likely not the driving force behind the CDW order in FeGe because long-range order is absent in FG560.

Next, we discuss the CDW order itself. We have clearly observed the consequences of the development of the long-range CDW order in FG320 on the electronic structure and their absence in FG560. Such evidence of the CDW order is observed on the FSs, the appearance of additional folded bands due to in-plane superlattice, the opening of a CDW gap on VHS1, as well as clear band shifts across $T_{CDW}$ in FG320 observed via a continuous temperature evolution measurement. The contrasting behavior in FG560 of no band shift nor band broadening near 110 K indicates that the lack of CDW order is not due to spatial inhomogeneity of the sample but truly a lack of ordering. This demonstrates the effectiveness of the thermal annealing procedure for tuning the CDW order in FeGe.

Finally, we comment on the lowest transition observed at $T^*$ in FG320C. Below $T_{CDW}$, as the only known transition in kagome FeGe is the spin canting transition, we first consider whether the observed band evolution at $T^*$ is associated with the spin canting. From a simple density functional theory calculation, the rotation of the spin away from *c*-axis due to canting does not change the electronic structure to produce observable effects (see Fig. S3). Furthermore, as the band shift at $T^*$ is only observed in FG320C and not in FG560C, this is in contrast to the fact that spin canting is observed in both types of annealed crystals. Therefore, we conclude that this band shift at $T^*$ is not associated directly with the spin canting. Recently, inelastic neutron measurements of the low energy magnetic excitations in as-grown crystals have revealed the competition of both itinerant and local magnetism associated with the canting transition, which may suggest contributions by the electronic degree of freedom to a potential spin density wave order formation below CDW order[31]. It remains to be seen how this spin density wave order is affected by the annealing process. More interestingly, we note that a lattice symmetry change has been reported by Raman measurements on as-grown samples [32]. The lattice symmetry was observed to lower from hexagonal to monoclinic upon entering the AFM order at $T_N$, and the symmetry is recovered to the higher symmetry at a temperature scale lower than $T_{CDW}$. This symmetry reversal behavior is qualitatively similar to our observed band evolution associated with the $\overline{\text{K}}$ point, which reverses between $T_\text{CDW}$ and $T^*$. This band shift was interpreted to be associated with the Ge dimerization[38]. It is interesting to point out that the non-monotonic behavior of this band shift is qualitatively distinct from the order parameter of the CDW order as measured by neutron diffraction peak[26] or the CDW gap size (Fig. 3D), and may suggest an intricate coupling of lattice, charge, and magnetism.

Taking all observations together in combination with studies in the literature, we come to the understanding that while thermal annealing does not drastically change the lattice, magnetism or the basic electronic structure, its most dominant effect is likely to be a tuning of the level of lattice disorder, which given all other properties being the same, ultimately suppresses the long-range CDW order. There are suggestions that the CDW is due to dimerization of Ge atoms in the kagome layer, and random distribution of dimerized Ge can suppress a long-range CDW with persistent short-range fluctuation[35,37,39]. With that said, our measurements also indicate a strong coupling of the lattice, spin and electronic degrees of freedom in kagome FeGe, which conspire together to give rise to the CDW order in this kagome magnet.

## Method

### Single crystal synthesis and post annealing

Single crystals of FeGe were synthesized following the reported recipe using chemical vapor transport [26,28]. We chose 560°C and 320°C as the annealing temperatures for our post-growth annealing protocols, same as the previous reports [36]. The as-grown crystals were sealed in an evacuated quartz tube and placed in a box furnace and set the temperature to 320°C. After maintaining this temperature for 96 hours, subsequently, the quartz tube was promptly removed from the furnace and quenched in tap water to room temperature. The same process was repeated for annealing the as-grown FeGe single crystals at 560°C. The resulting annealed crystals maintain their original lustrous appearance, showing no discernible surface changes due to the annealing process.

### Magnetic susceptibility measurement

The magnetization as a function of temperature and field was measured by using a Quantum Design DynaCool Physical Properties Measurement System (PPMS-9T) over the temperature range from 2 K to 300 K and under 0.1 T magnetic field. Post annealed FG320 and FG560 single crystals were separately mounted on PPMS sample holders, having the crystallographic *c* axis perpendicular to the 0.1T field. The orientation of the crystal was determined by standard Laue techniques before mounting to PPMS sample holder.

**Angle-resolved photoemission spectroscopy**

ARPES measurements were carried out at BL5-2 of the Stanford Synchrotron Radiation Lightsource with a DA30 electron analyzer. Post-annealed FeGe single crystals were cleaved to have normal surfaces face the (001) direction in ultra-high vacuum environment with a base pressure better than $5\times10^{-11}$ torr. We have identified two different types of termination layer as our previous study did and focused on the Ge termination to trace the features from the CDW order more evidently [28]. Energy and angular resolutions were better than 10 meV and 0.1°, respectively. We have adapted beam spot smaller than $50 \times 50$ μm$^2$. $E_F$ is corrected by a measurement of freshly deposited gold near crystals, and both are electrically connected, guaranteeing to have the same $E_F$.

**Density Functional Theory**

Electronic structures of FeGe are calculated using density functional theory as implemented in the Vienna Ab-initio Simulation Package [40]. The crystal structure of FeGe is fully relaxed under the antiferromagnetic configuration until the maximal remaining force is smaller than one meV/Å. Spin-orbit coupling is not considered in structure relaxation. The Perdew-Burke-Ernzerhof-type generalized gradient approximation [41] mimics the electron-electron exchange interaction throughout. The energy cutoff for the plane wave basis set is 350 eV. Brillouin zones of the antiferromagnetic and spin canting phases are sampled with a *k*-mesh of 12×12×8. Band structures are obtained with the spin-orbit coupling.

**Acknowledgments**

Use of the Stanford Synchrotron Radiation Lightsource, SLAC National Accelerator Laboratory, is supported by the U.S. Department of Energy, Office of Science, Office of Basic Energy Sciences under Contract No. DE-AC02-76SF00515. The ARPES work at Rice University was supported by the U.S. DOE grant No. DE-SC0021421, the Gordon and Betty Moore Foundation's EPiQS Initiative through grant No. GBMF9470 and the Robert A. Welch Foundation Grant No. C-2175 (MY). The single-crystal synthesis work at Rice was supported by the U.S. DOE, BES under Grant No. DE-SC0012311 (PD). Work at the University of California, Berkeley and Lawrence Berkeley National Laboratory was funded by the U.S. DOE, Office of Science, Office of Basic Energy Sciences, Materials Sciences and Engineering Division under Contract No. DE-AC02-05CH11231 (Quantum Materials Program KC2202) (RJB). We acknowledge the support from National Science Foundation (NSF) grants Nos. DMR-1921798 and DMR-2324032 (JSO, RJB, MY).

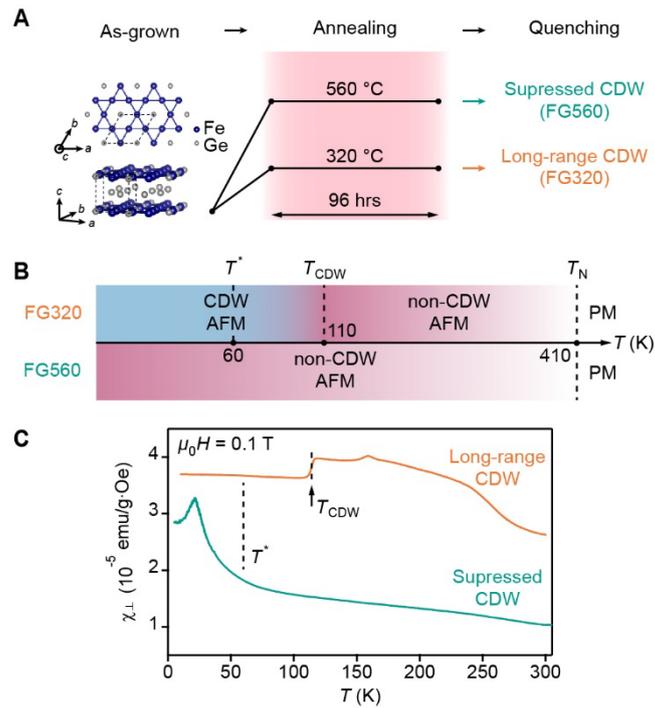

**Fig. 1. Characterizations for long-range and suppressed CDW order.** (**A**) Crystal structure of FeGe from top and side perspectives. Post-annealing conditions for achieving long-range (FG320) or suppressed (FG560) CDW order. (**B**) Schematic illustration for the distinct phases in FG320 and FG560 as a function of temperature. (**C**) Magnetic susceptibility with field perpendicular to the (001) direction.

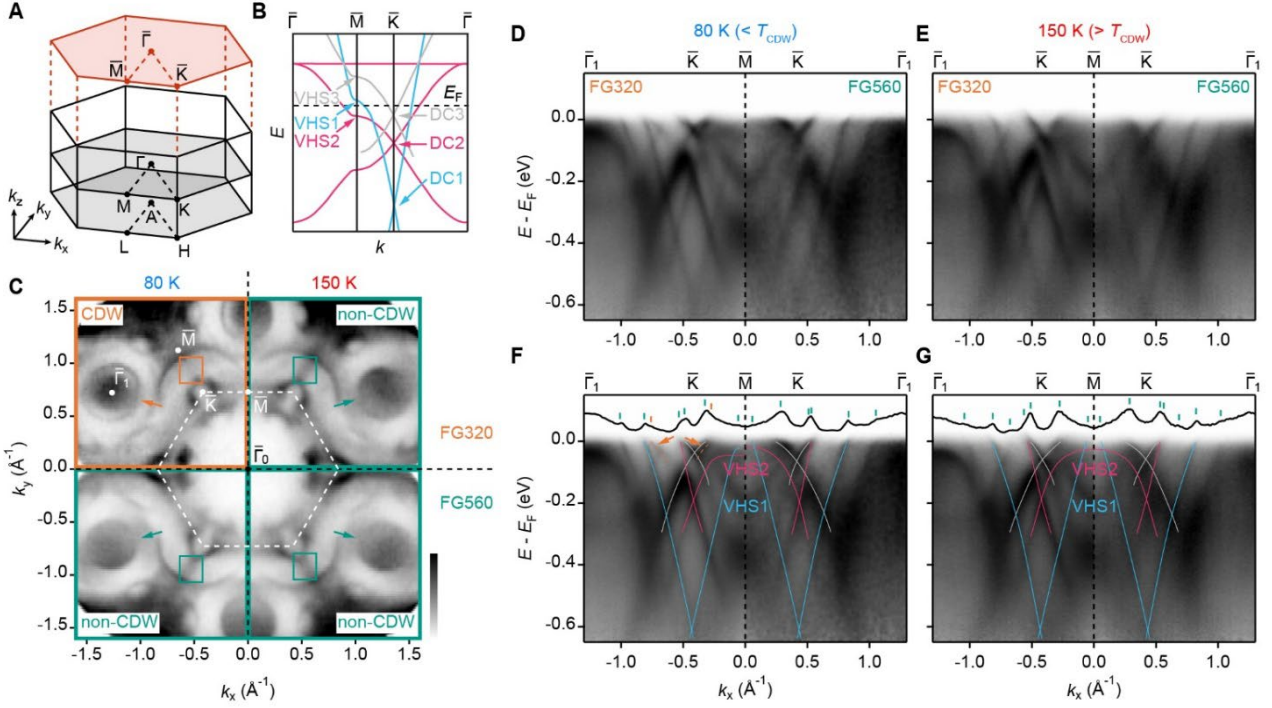

**Fig. 2. Comparison of electronic structures between FG320 and FG560 below and above $T_{CDW}$.** (**A**) Brillouin zone and locations of high symmetry points of FeGe. Surface-projected Brillouin zone is also depicted. (**B**) Schematic band dispersions for FeGe. (**C**) FS maps obtained from FG320 (top) and FG560 (bottom) at 80 K (<$T_{CDW}$; left) and 150 K (>$T_{CDW}$; right), as plotted in the four quadrants. Orange and green arrows/boxes mark distinct features when comparing CDW ordered and normal phases. (**D**) A side-by-side comparison of band dispersions along the $\overline{K}$-$\overline{M}$-$\overline{K}$ between FG320 and FG560 at 80 K. (**E**) Similar comparison at 150 K. (**F**) Same plot as (**D**) with guides for the eye to trace VHS1 and VHS2 band dispersions. MDC at $E_F$ is overlaid showing peak locations of $E_F$-crossing band dispersions. We mark peak locations with green bars and differentiate extra bands in the CDW phase with orange bars. (**G**) Same plot as (**F**) but for 150 K. Note that no extra band crossing at $E_F$ is observed.

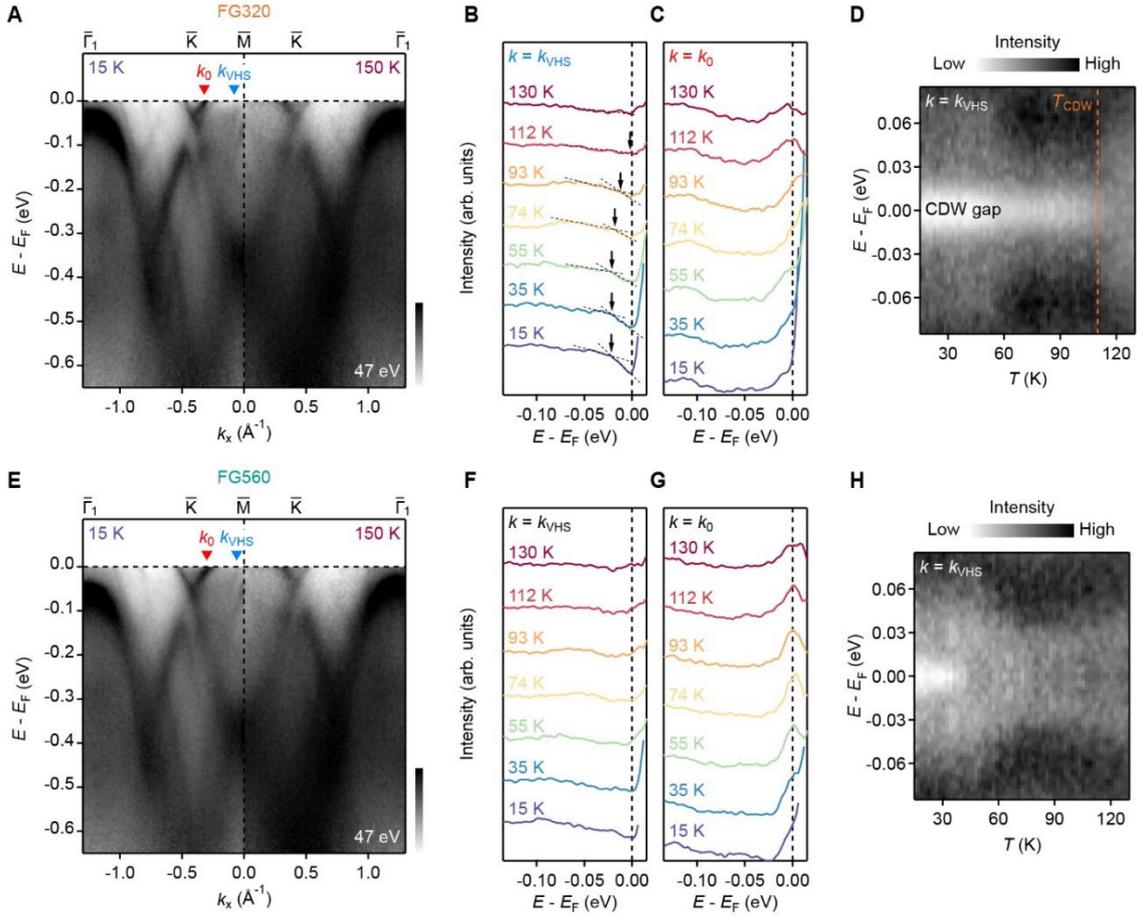

**Fig. 3. CDW gap in FG320 and FG560.** (**A**) A side-by-side comparison of band dispersions from FG320 along the $\overline{\text{K}}$-$\overline{\text{M}}$-$\overline{\text{K}}$ taken at 80 K (left) and 150 K (right). Two representative Fermi momenta are annotated as $k_{\text{VHS}}$ and $k_0$. (**B**) EDCs at various temperatures obtained at $k_{\text{VHS}}$ as marked in (**A**). We divided out the Fermi-Dirac distributions from all the EDCs, revealing a dip feature near $E_F$ below $T_{\text{CDW}}$ (= 110 K). Arrows point at where the dashed guidelines cross, which indicates a slope change in the EDCs as gap opens. (**C**) Same plot as (**B**) obtained at $k_0$. Contrary to (**B**), we observe peaks at $E_F$ at all temperatures. (**D**) Stacked plots of symmetrized EDCs at $k_{\text{VHS}}$. The white region shows where spectral intensity is suppressed, corresponding to the CDW gap region. (**E**–**H**) Analogous plots to (**A**–**D**) from FG560. Note that CDW gap is not as evident as it is in FG320. Smaller white region in (**H**) than that in (**D**) is from suppressed CDW order.

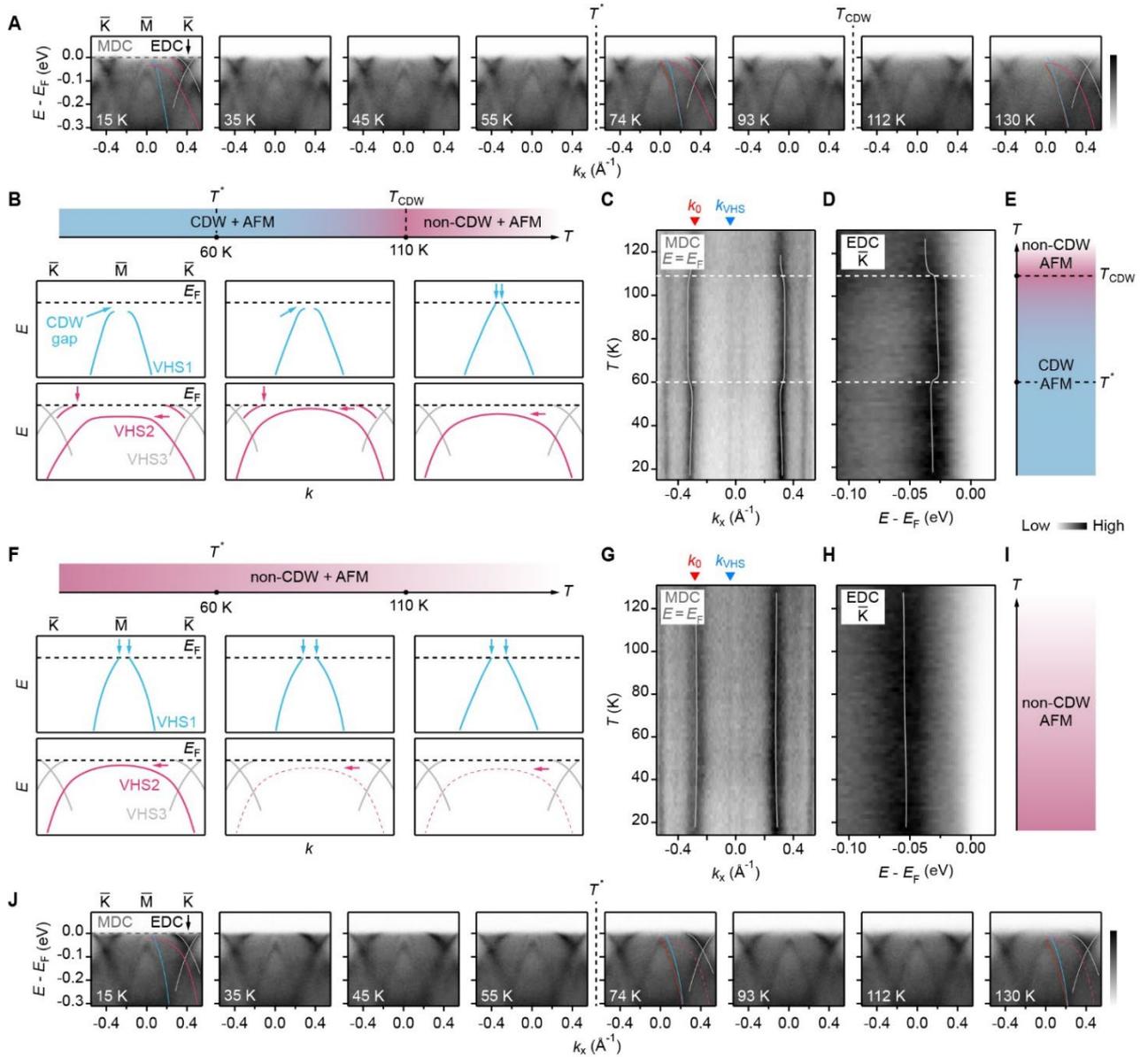

**Fig. 4. Band dispersions at representative temperatures and their evolutions in FG320 and FG560.** (**A**) High symmetry cuts along the $\overline{K}$-$\overline{M}$-$\overline{K}$ at representative temperatures from FG320. $T^*$ and $T_{CDW}$ are indicated. MDC (grey dashed line) and EDC (black arrow at $k_x = 0$) are locations where we take a stack MDCs shown in (**C**) and two stacks of EDCs in (**D**). (**B**) Schematics showing how band dispersions evolve as a function of temperature. Notable energy and momentum locations are pointed by colored arrows. Phase indicator bar is put together above the schematics, matching the horizontal temperature scale. (**C**) A stacked MDC at $E_F$, showing non-monotonic temperature evolutions of Fermi momenta across $T^*$ (= 60 K) and $T_{CDW}$ (= 110 K). (**D**) Stacked EDC at $\overline{K}$, which present the band shift across $T_{CDW}$ and a reverted behavior at $T^*$ (**E**) Phase indicator whose temperature axis is aligned with (**C**–**D**). (**F**–**I**) Analogous plots to (**B**–**E**) from FG560. (**J**) High symmetry cuts along $\overline{M}$ and $\overline{K}$ at representative temperatures from FG560.